\documentclass[aps,prc,reprint,notitlepage,floatfix,superscriptaddress,nofootinbib]{revtex4-1}
\usepackage[utf8]{inputenc}
\usepackage{amsmath}
\usepackage{amssymb}
\usepackage{graphicx}

\usepackage{mathptmx}
\usepackage{bm}

\usepackage{xcolor}
\usepackage[colorlinks,breaklinks]{hyperref}
\hypersetup{linkcolor=[HTML]{0000DD},
            citecolor=[HTML]{DC143C},
            urlcolor=[HTML]{00008B}}

\usepackage{multirow}
\usepackage{dcolumn}

\newcommand{\hatx}{\hat{\mathbf{x}}}
\newcommand{\haty}{\hat{\mathbf{y}}}
\newcommand{\hatz}{\hat{\mathbf{z}}}

\newcommand{\hatX}{\hat{\mathbf{X}}}
\newcommand{\hatY}{\hat{\mathbf{Y}}}
\newcommand{\hatZ}{\hat{\mathbf{Z}}}

\newcommand{\VPP}{\mathbf{P}_{P}}
\newcommand{\VPD}{\mathbf{P}_{D}}
\newcommand{\hatPP}{\hat{\mathbf{P}}_{P}}

\newcommand{\ps}{\mathbf{p}^{*}}
\newcommand{\hatps}{\hat{\mathbf{p}}^{*}}

\newcommand{\rhoiM}{\rho^{i}_{M_{P};M_{P}^{\prime}}}
\newcommand{\rhofDX}{\rho^{f}_{\la_{D}\la_{X};\la_{D}^{\prime}\la_{X}^{\prime}}}
\newcommand{\rhoDD}{\rho^{D}_{\la_{D};\la_{D}^{\prime}}}

\newcommand{\Tch}{T_{\mathrm{ch}}}
\newcommand{\Tkin}{T_{\mathrm{kin}}}
\newcommand{\sNN}{\sqrt{s_{\mathrm{NN}}}}
\newcommand{\pT}{p_{\mathrm{T}}}
\newcommand{\mT}{m_{\mathrm{T}}}

\newcommand{\al}{\alpha}
\newcommand{\be}{\beta}
\newcommand{\ga}{\gamma}

\newcommand{\de}{\delta}
\newcommand{\De}{\Delta}

\newcommand{\ths}{\theta^{*}}
\newcommand{\phs}{\phi^{*}}

\newcommand{\la}{\lambda}
\newcommand{\La}{\Lambda}
\newcommand{\Si}{\Sigma}
\newcommand{\Om}{\Omega}

\renewcommand{\l}{\left}
\renewcommand{\r}{\right}
  \newcommand{\<}{\langle}
\renewcommand{\>}{\rangle}

\newcommand{\diag}{\mathrm{diag}}
\newcommand{\tr}{\mathrm{tr}}
\newcommand{\re}{\mathrm{Re}}
\newcommand{\im}{\mathrm{Im}}
\newcommand{\pd}{\partial}

\begin{document}

\title{Feed-down effect on $\Lambda$ spin polarization}
\author{Xiao-Liang Xia}\email{xiaxl@fudan.edu.cn}
\affiliation{Department of Physics, Center for Particle Physics and Field Theory, Fudan University, Shanghai 200433, China}
\author{Hui Li}\email{lihui$_$fd@fudan.edu.cn}
\affiliation{Key Laboratory of Nuclear Physics and Ion-beam Application (MOE), Fudan University, Shanghai 200433, China}
\author{Xu-Guang Huang}\email{huangxuguang@fudan.edu.cn}
\affiliation{Department of Physics, Center for Particle Physics and Field Theory, Fudan University, Shanghai 200433, China}
\affiliation{Key Laboratory of Nuclear Physics and Ion-beam Application (MOE), Fudan University, Shanghai 200433, China}
\author{Huan Zhong Huang}\email{huanzhonghuang@fudan.edu.cn}
\affiliation{Key Laboratory of Nuclear Physics and Ion-beam Application (MOE), Fudan University, Shanghai 200433, China}
\affiliation{Department of Physics and Astronomy, University of California, Los Angeles, California 90095, USA}

\begin{abstract}
We develop a theoretical framework to study the feed-down effect of higher-lying strange baryons on the spin polarization of the $\Lambda$ hyperon. In this framework, we consider two-body decays through strong, electromagnetic, and weak processes and derive general formulas for the angular distribution and spin polarization of the daughter particle by adopting the helicity formalism. Using the realistic experimental data as input, we explore the feed-down contribution to the global and the local $\Lambda$ polarizations and find that such a contribution suppresses the primordial $\Lambda$ polarization, which is not strong enough to resolve the discrepancy between the current theoretical and the experimental results on the azimuthal-angle dependence of $\Lambda$ polarization. Our paper may also be useful for the measurement of spin polarization of baryons heavier than $\Lambda$ (e.g., $\Xi^-$) in future experiments.
\end{abstract}
\maketitle

\section{Introduction}

The recent measurement of the spin polarization of $\La$ and $\bar{\La}$ hyperons (hereafter, ``$\La$ polarization'' for simplicity) provided strong evidence for the existence of large fluid vorticity in the hot and dense matter created in noncentral relativistic heavy ion collisions~\cite{STAR:2017ckg}. This finding, for the first time, showed a physical connection between the fireball's vorticity and the spin polarization of the final-state hadrons and opened the door to the study of various phenomena in the presence of vorticity or rotation in the strongly interacting quark-gluon plasma. Such phenomena include, in addition to the spin polarization of baryons~\cite{Liang:2004ph,Voloshin:2004ha,Gao:2007bc,Becattini:2007sr,Huang:2011ru}, the spin alignment of vector mesons~\cite{Liang:2004xn,Abelev:2008ag,Zhou:2019lun,Singh:2018uad}, the chiral vortical effect or wave~\cite{Erdmenger:2008rm,Banerjee:2008th,Son:2009tf,Liu:2018xip,Jiang:2015cva}, the emergence of spin transport coefficients~\cite{Hattori:2019lfp}, the dissociation of chiral condensate~\cite{Chen:2015hfc,Ebihara:2016fwa,Chernodub:2016kxh,Chernodub:2017ref,Wang:2018zrn,Wang:2019nhd}, and the modifications of the QCD phase diagram~\cite{Jiang:2016wvv,Huang:2017pqe,Liu:2017spl,Wang:2018sur,Zhang:2018ome}.

The measurement in Ref.~\cite{STAR:2017ckg} is for the mean value of the $\La$ polarization in the midrapidity region (dubbed the \textit{global polarization}) which reflects the space-averaged value of the vorticity. Such a space-averaged vorticity, in turn, reflects the global angular momentum of the colliding system. In addition to this, more detailed measurements were performed very recently~\cite{Adam:2018ivw,Adam:2019srw,Niida:2018hfw}, exhibiting $\La$ polarization as a function of the transverse momentum, azimuthal angle, and rapidity. These new measurements indicated that the vorticity field, if assumed to be responsible for the detailed structure in the $\La$ polarization, may have a very nontrivial local structure in the fireball. Indeed, it has been proposed from model simulations that the vorticity can be generated from different sources, leading to a novel local vortical structure and local polarization~\cite{Betz:2007kg,Baznat:2013zx,Baznat:2015eca,Becattini:2013vja,Csernai:2013bqa,Csernai:2014ywa,Becattini:2015ska,Teryaev:2015gxa,Jiang:2016woz,Deng:2016gyh,Pang:2016igs,Karpenko:2016jyx,Xie:2016fjj,Li:2017slc,Shi:2017wpk,Becattini:2017gcx,Xia:2018tes,Wei:2018zfb,Sun:2018bjl}. The total $\La$ polarization is the superposition of them. Particularly, anisotropic flow on the transverse plane can produce a quadrupole pattern of the longitudinal vorticity component, and accordingly, there is a \textit{longitudinal local polarization} $P_{z}$ where the $z$ axis is along the beam direction~\cite{Becattini:2017gcx,Voloshin:2017kqp,Xia:2018tes}. Similarly, the nonuniform transverse expansion of the fireball along the longitudinal axis can produce transverse vorticity circling the $z$ axis from which the \textit{transverse local polarization} $(P_{x},P_{y})$ can be generated where the $x$ axis is along the impact parameter and the $y$ axis is perpendicular to the reaction plane~\cite{Xia:2018tes,Wei:2018zfb}. Besides the circling polarization, it is also found that the polarization $P_y$ at midrapidity has a difference from the in-plane direction to the out-of-plane direction~\cite{Niida:2018hfw,Karpenko:2016jyx,Wei:2018zfb}.

However, there are discrepancies between experimental measurements of the $\La$ polarization and theoretical calculations, in particular, the predicted azimuthal-angle dependence of the longitudinal and transverse spin polarizations at midrapidity has the opposite sign compared to the data~\cite{Adam:2019srw,Niida:2018hfw}. This constitutes a remarkable puzzle and challenges the thermal-vorticity interpretation of the $\La$ polarization which assumes that the $\La$ polarization is simply proportional to the thermal vorticity~\cite{Becattini:2013fla,Becattini:2016gvu,Fang:2016vpj}.

To resolve this puzzle, one important issue should be understood first, that is, the feed-down contributions from decays of other strange baryons to the final $\La$ polarization. This is because only a fraction of the final-state $\La$ and $\bar{\La}$ hyperons are produced directly at the hadronization stage (which will be called the \textit{primordial} $\La$ and $\bar{\La}$ and their spin polarizations may reflect the information of the vorticity). A big fraction of $\La$ and $\bar{\La}$ hyperons are from the decays of higher-lying strange baryons, such as $\Sigma^0$, $\Sigma^*$, $\Xi$, etc. Thus, to bridge the measured $\La$ polarization and the information of the vorticity, we must take into account the correction from the feed-down contributions. As will be shown in Sec.~\ref{sec2}, $\La$ hyperons produced by particle decay may have an anisotropic angular distribution in the rest frame of the parent particle, and the $\La$ polarization vector depends on its emitted direction. These effects and the interplay between them can re-distribute the $\La$ polarization in azimuthal-angle space and provide a possible solution to the above puzzle.

The purpose of this paper is to systematically investigate the feed-down effect on the $\La$ polarization. We stress that, for the global polarization case, this problem has been studied in Ref.~\cite{Becattini:2016gvu} where the effect of feed-downs was found to give a linear relation between $\langle\mathbf{P}_{D}\rangle$ and $\mathbf{P}_{P}$, that is,
\begin{equation}
\langle\mathbf{P}_{D}\rangle=C\mathbf{P}_{P},\label{eq:linear rule}
\end{equation}
where $\langle\mathbf{P}_{D}\rangle$ is the momentum-averaged (i.e., the global) polarization of the daughter particle $D$ and $\mathbf{P}_{P}$ is the polarization vector of the parent particle. The coefficient $C$ is the polarization transfer factor between the parent and the daughter particle. However, Eq.~(\ref{eq:linear rule}) cannot be used to study the effect of particle decay on the local $\La$ polarization which we will focus on.

This paper is organized as follows. In Sec.~\ref{sec2}, we will derive a set of formulas for the angular distribution and polarization vector of the daughter particle produced in a two-body decay. Several different decay channels will be considered. Based on these formulas, we will implement the numerical simulation to study the effect of feed-downs on $\La$ polarization. The numerical results and discussions will be presented in Sec.~\ref{sec3}. Finally, a summary will be given in Sec.~\ref{sec4}.

\section{Spin polarization in two-body decay} \label{sec2}

To study the effect of particle decay on the $\La$ polarization, we need to first answer two questions: (1) For a given decay channel with the polarization vector of the parent particle given as $\VPP$, what is the angular distribution of the specified daughter particle (i.e., $\La$ hyperon in our case), and (2) for a daughter particle emitted along a specific direction $\hatps$ (a hat over a vector denotes the unit vector) in the parent's rest frame, what is its polarization vector $\VPD$?

As the most important decay channels to produce $\La$ are two-body decays, let us consider a generic two-body decay process,
\begin{equation}
P\to D+X,
\end{equation}
where $P$ is the parent and $D$ and $X$ are the daughters. Among them, $D$ stands for the $\La$ hyperon or a particle that can further decay to $\La$, whereas $X$ is a by-product particle. The goal in this section is to find the angular distribution of $D$ and calculate its polarization vector $\VPD$ as a function of its emission direction in the rest frame of $P$ with the polarization vector $\VPP$ of $P$ fixed.

This problem for some decay channels was already studied in the 1950s, for example, the weak decay of spin-1/2 hyperons in Refs.~\cite{Lee:1957he,Lee:1957qs} and the electromagnetic (EM) decay process $\Si^{0}\to\La\ga$ in Ref.~\cite{Gatto:1958qmn}. Later, a systematic method, called the helicity formalism, was established to study the spin-related problem in 1959~\cite{Jacob:1959at}. By using this method, we can easily deal with the spin-polarization problem for all the decay channels that are needed in this paper. In this section, we will first introduce the basic framework of the helicity formalism, and then, we will apply this formalism to some specific decay channels. For more information about the helicity formalism, we refer the readers to Refs.~\cite{Chung:1971ri,Richman:1984gh,devanathan1999angular}.

\begin{figure}[h]
\centering \includegraphics[width=0.6\columnwidth]{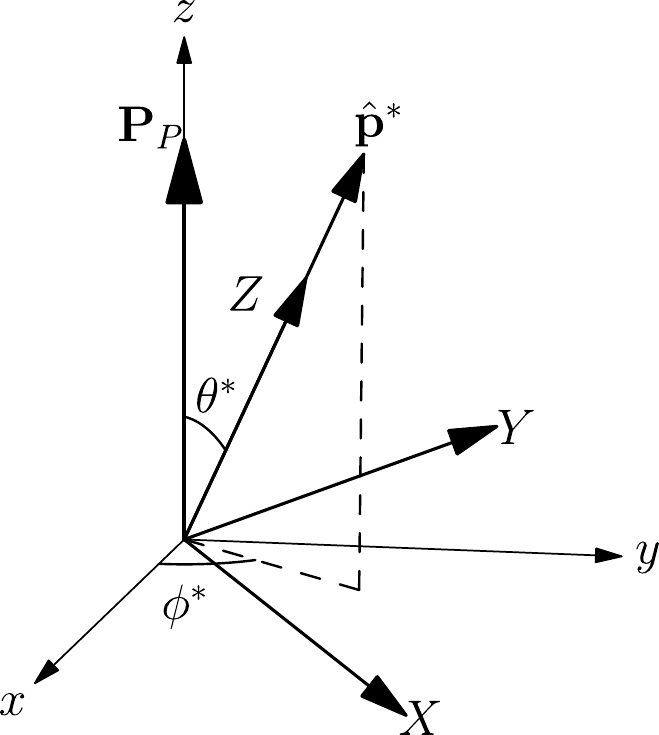}
\caption{Illustration of the coordinate frames in helicity formalism.}
\label{fig:frame}
\end{figure}

Figure~\ref{fig:frame} illustrates the coordinate frames which are used in the  helicity formalism. For the decay problem, it is natural to use the rest frame of the parent $P$ (RFP). The $z$ axis is set to be along the polarization vector of $P$, i.e., $\hatz=\hatPP$. The transverse axes $\hatx$ and $\haty$ can be arbitrarily chosen since the whole system has a rotational symmetry around the $z$ axis. In this $x$-$y$-$z$ frame, we can label the spin state of $P$ as
\begin{equation}
|i\>=|S_{P}M_{P}\>,
\end{equation}
where $S_{P}$ is the spin number of $P$ and $M_{P}$ is its spin projection along the $z$ axis.

After the decay, the momenta of $D$ and $X$ are equal but along opposite directions in the RFP. We denote the momentum of $D$ in the RFP as $\ps$, whose polar and azimuthal angles are $\ths$ and $\phs$, respectively. In the helicity formalism, the spin states of $D$ and $X$ are quantized along the direction of $\ps$ (i.e., the spin states are described by the helicities of $D$ and $X$). Accordingly, we introduce new frame axes $(\hatX,\hatY,\hatZ)$ as illustrated in Fig.~\ref{fig:frame}, where $\hatZ=\hatps$ is the new quantization direction and $\hatX$ and $\hatY$ directions could be arbitrarily chosen. However, it would be convenient if we fix the choice by taking $\hatY$ being along $\hatz\times\hatZ$ and $\hatX$ being along $\hatY\times\hatZ$. This follows the convention in Refs.~\cite{Chung:1971ri,devanathan1999angular}. The explicit forms of the new frame basis are
\begin{align}
\hatX & =\frac{\l(\hatPP\times\hatps\r)\times\hatps}{\sin\ths}=\frac{\l(\hatPP\cdot\hatps\r)\hatps-\hatPP}{\sin\ths},\label{eq:new-X}\\
\hatY & =\frac{\hatPP\times\hatps}{\sin\ths},\label{eq:new-Y}\\
\hatZ & =\hatps,\label{eq:new-Z}
\end{align}
where $\sin\ths$ in the denominators makes $\hatX$ and $\hatY$ unit vectors. This choice of the $X$-$Y$-$Z$ frame can be achieved from the $x$-$y$-$z$ frame by a standard rotating operation $\widehat{R}(\phi,\theta,\varphi)$ with Euler angles $\phi=\phs$, $\theta=\ths$, and $\varphi=0$~\cite{Chung:1971ri,devanathan1999angular}.

In the $X$-$Y$-$Z$ frame, the spin states of $D$ and $X$ can be labeled as $|S_{D}\la_{D}\>$ and $|S_{X}\la_{X}\>$ respectively, where $S_{D}$ and $S_{X}$ are their spin numbers and $\la_{D}$ and $\la_{X}$ are their helicities. The spin projections along the $Z$ axis of $D$ and $X$ are $\la_{D}$ and $-\la_{X}$, respectively. For massive particles, $\la_{D}=-S_{D},\dots,S_{D}$ and $\la_{X}=-S_{X},\dots,S_{X}$. If $X$ is a massless particle, such as a photon, $\la_{X}$ takes binary values $\pm S_{X}$.

As we are concerned about the angular distribution of $D$ and its spin state varying with the momentum direction, we further define the following final state (the magnitude of ${\bf p}^*$ is not expressed),
\begin{equation}
|f\>=|\ths\phs\la_{D}\la_{X}\>.
\end{equation}
It labels a pair of $D$ and $X$ being emitted along the $\hatps$ and $-\hatps$ directions with their helicities being $\la_{D}$ and $\la_{X}$, respectively.

After the initial and final states are defined, it comes to a useful result of the helicity formalism. Saying the decay process is encoded by a decay operator $\widehat{H}$, the spin-density matrix of the final-state $\rho^{f}$ is related to that of the initial-state $\rho^{i}$ by
\begin{equation}
\rhofDX(\ths,\phs)=\sum_{M_{P},M_{P}^{\prime}}H_{\la_{D}\la_{X};M_{P}}\rhoiM H_{M_{P}^{\prime};\la_{D}^{\prime}\la_{X}^{\prime}}^{\dagger},\label{eq:density matrix}
\end{equation}
where $\rhoiM$ and $\rhofDX$ are the matrix elements of $\rho^{i}$ and $\rho^{f}$, respectively. They are defined as
\begin{align}
\rhoiM & =\<S_{P}M_{P}|\rho^{i}|S_{P}M_{P}^{\prime}\>,\\
\rhofDX(\ths,\phs) & =\<\ths\phs\la_{D}\la_{X}|\rho^{f}|\ths\phs\la_{D}^{\prime}\la_{X}^{\prime}\>.
\end{align}
Note that $\rho^{f}$ is labeled by $\la_{D}$ and $\la_{X}$ jointly and is a function of $\ths$ and $\phs$. The normalization of $\rho^{i}$ and $\rho^{f}$ are, respectively,
\begin{align}
\tr(\rho^{i}) & =1,\\
\int d\Om^{*}\tr(\rho^{f}) & =1,
\end{align}
with $d\Om^{*}=d\cos\ths d\phs$ being the solid angle volume and $\tr$ being the trace over the spin states.

According to Refs.~\cite{Chung:1971ri,Richman:1984gh,devanathan1999angular}, the matrix element $H_{\la_{D}\la_{X};M_{P}}$ in Eq.~(\ref{eq:density matrix}) is given by
\begin{equation}
H_{\la_{D}\la_{X};M_{P}}=\sqrt{\frac{2S_{P}+1}{4\pi}}D_{M_{P};\la_{D}-\la_{X}}^{S_{P}*}(\phs,\ths,0)A_{\la_{D};\la_{X}},\label{eq:h matrix}
\end{equation}
where $D_{M_{P};\la_{D}-\la_{X}}^{S_{P}}$ is the Wigner $D$ function and $D_{M_{P};\la_{D}-\la_{X}}^{S_{P}*}$ is its conjugate. The arguments $(\phs,\ths,0)$ correspond to the Euler angles for our choice of $X$-$Y$-$Z$ axes. $A_{\la_{D};\la_{X}}$ is the relative dynamical amplitude for the decay process from $|S_{P}M_{P}\>$ to $|\la_{D}\la_{X}\>$, and its value depends only on $\la_{D}$ and $\la_{X}$ and is normalized as $\sum_{\la_{D},\la_{X}}|A_{\la_{D};\la_{X}}|^{2}=1$.

For the current paper, it is not necessary to calculate the exact value of $A_{\la_{D};\la_{X}}$. Instead, one can use some constraining conditions to simplify the calculation. If the decay process is parity conserved (strong and EM decays), $A_{\la_{D};\la_{X}}$ is constrained by the following relation~\cite{Chung:1971ri,Richman:1984gh,devanathan1999angular}:
\begin{equation}
A_{\la_{D};\la_{X}}=\pi_{P}\pi_{D}\pi_{X}(-1)^{S_{P}-S_{D}-S_{X}}A_{-\la_{D};-\la_{X}},\label{eq:parity conserve}
\end{equation}
where $\pi_{P}$, $\pi_{D}$, and $\pi_{X}$ are the parity values ($\pm1$) of $P$, $D$, and $X$, respectively. In this case, we always have $|A_{\la_{D};\la_{X}}|=|A_{-\la_{D};-\la_{X}}|$. On the other hand, for parity-violating weak decay, Eq.~(\ref{eq:parity conserve}) does not apply, and $A_{\la_{D};\la_{X}}$ can be parametrized by several decay parameters; for example, $1/2\to 1/2\ 0$ weak decay can be parametrized by three decay parameters $\al$, $\be$, and $\ga$ (see below and Ref.~\cite{Lee:1957qs}) whose values can be determined by fitting the experimental measurements or by concrete field-theory calculations.

Up to this point, the main line of the helicity formalism has been established. From a given initial spin-density-matrix $\rho^{i}$, we can calculate the spin-density matrix of final-state $\rho^{f}$ by Eq.~(\ref{eq:density matrix}). Then, the angular distribution of $D$ in the RFP can be determined by
\begin{equation}
\frac{1}{N}\frac{dN}{d\Om^{*}}=\tr\l(\rhofDX\r).\label{eq:dN/dOmega define}
\end{equation}
To obtain the polarization vector of $D$ emitted in a certain direction, we first take the partial trace of $\rho^{f}$ over index $\la_{X}$, obtaining the spin-density matrix of $D$,
\begin{equation}
\rhoDD=\tr_{X}\l(\rhofDX\r),
\end{equation}
then the polarization vector of $D$ can be calculated by
\begin{equation}
\VPD=\tr_{D}\l(\widehat{\mathbf{P}}\rhoDD\r)/\tr_{D}\l(\rhoDD\r),\label{eq:P_D define}
\end{equation}
where $\widehat{\mathbf{P}}$ is the polarization operator, e.g., $\widehat{\mathbf{P}}=\bm{\sigma}$ (the Pauli matrix) for spin-1/2 particles. As the spin-density-matrices $\rhofDX$ and $\rhoDD$ are functions of $\ths$ and $\phs$, the polarization vector $\VPD$ in Eq.~(\ref{eq:P_D define}) depends on the momentum direction $\hatps$ in general.

Now, we apply the above results to the decay channels relevant for $\La$ production to calculate the angular distribution of $D$ in the RFP and its polarization vector $\VPD$.

\subsection{Strong decay $1/2^{\pm}\to1/2^{+}0^{-}$}

We start with the simplest case, the strong decay $1/2^{\pm}\to 1/2^{+}0^{-}$. As we choose the polarization direction of $P$ to be the initial spin-quantization axis, the spin-density matrix of $P$ is a diagonal matrix, which reads
\begin{equation}
\rhoiM=\diag\l(\frac{1+P_{P}}{2},\frac{1-P_{P}}{2}\r),\label{eq:rho^i 1/2}
\end{equation}
where $P_{P}$ is the polarization magnitude of $P$.

Using Eqs.~(\ref{eq:density matrix}) and (\ref{eq:h matrix}), we obtain the final spin-density matrix as
\begin{widetext}
\begin{equation}
\rhoDD=\frac{1}{4\pi}\begin{pmatrix}
|A_{1/2}|^{2}\l(1+P_{P}\cos\ths\r) & -A_{1/2}A_{-1/2}^{*}\,P_{P}\sin\ths\\
-A_{1/2}^{*}A_{-1/2}\,P_{P}\sin\ths & |A_{-1/2}|^{2}\l(1-P_{P}\cos\ths\r)
\end{pmatrix},\label{eq:rho^D 1/2}
\end{equation}
\end{widetext}
where $\la_{D}$ takes binary values $\la_{D}=\pm1/2$ and the index of $A$ refers to $\la_{D}$. We have omitted the index $\la_{X}$ as it takes a single value $\la_{X}=0$, i.e., $A_{\la_D}\equiv A_{\la_D;0}$ and $\rhoDD\equiv\rho^f_{\la_D0;\la_D^{\prime}0}$.

Because strong decay conserves parity, the decay amplitude $A_{\pm1/2}$ is constrained by Eq.~(\ref{eq:parity conserve}). We have $A_{1/2}=-A_{-1/2}$ for decay channel $1/2^{+}\to1/2^{+}0^{-}$ and $A_{1/2}=+A_{-1/2}$ for channel $1/2^{-}\to1/2^{+}0^{-}$. For both channels, the normalization condition gives $|A_{1/2}|^{2}=|A_{-1/2}|^{2}=1/2$. Then Eq.~(\ref{eq:rho^D 1/2}) can be reduced to
\begin{equation}
\rhoDD=\frac{1}{8\pi}\begin{pmatrix}
1+P_{P}\cos\ths & \pm P_{P}\sin\ths\\
\pm P_{P}\sin\ths & 1-P_{P}\cos\ths
\end{pmatrix},\label{eq:rho^D 1/2 strong}
\end{equation}
where $\pm$ takes the upper sign for channel $1/2^{+}\to1/2^{+}0^{-}$ and the lower sign for $1/2^{-}\to1/2^{+}0^{-}$, respectively.

Plugging Eq.~(\ref{eq:rho^D 1/2 strong}) into Eqs.~(\ref{eq:dN/dOmega define}) and (\ref{eq:P_D define}), we find that the angular distribution of $D$ in the RFP is
\begin{equation}
\frac{1}{N}\frac{dN}{d\Om^{*}}=\frac{1}{4\pi},
\label{eq:strong1/2_angular}
\end{equation}
and the three components of $\VPD$ are
\begin{align}
P_{X}(\ths,\phs) & =\pm P_{P}\sin\ths,\\
P_{Y}(\ths,\phs) & =0,\\
P_{Z}(\ths,\phs) & =P_{P}\cos\ths.
\end{align}
Note that these three components are on the bases $\hatX$, $\hatY$, and $\hatZ$, so plugging them into $\VPD=P_{X}\hatX+P_{Y}\hatY+P_{Z}\hatZ$ and using Eqs.~(\ref{eq:new-X})--(\ref{eq:new-Z}), we can rewrite $\VPD$ as
\begin{equation}
\VPD=2\l(\VPP\cdot\hatps\r)\hatps-\VPP
\label{eq:strong 1/2+}
\end{equation}
for channel $1/2^{+}\to1/2^{+}0^{-}$ and
\begin{equation}
\VPD=\VPP
\label{eq:strong 1/2-}
\end{equation}
for channel $1/2^{-}\to1/2^{+}0^{-}$. Equations~(\ref{eq:strong 1/2+}) and (\ref{eq:strong 1/2-}) are expressed by $\VPP$ and $\hatps$, so they are free from the choice of the frames and are thus convenient for practical use.

\subsection{Weak decay $1/2\to1/2\ 0$}

Since the spin numbers for the weak decay $1/2\to1/2\ 0$ is the same as the strong decay above, the spin-density-matrix $\rhoDD$ for this channel is also given by Eq.~(\ref{eq:rho^D 1/2}). The difference is that $A_{\pm1/2}$ is no longer constrained by Eq.~(\ref{eq:parity conserve}). Indeed, the weak decay process $1/2\to1/2\ 0$ is a mixture of $s$-wave (parity-even) and $p$-wave (parity-odd) modes~\cite{Lee:1957he,Lee:1957qs}, so one can decompose $A_{\pm1/2}$ into
\begin{equation}
A_{\pm1/2}=\frac{A_{s}\pm A_{p}}{\sqrt{2\l(|A_{s}|^{2}+|A_{p}|^{2}\r)}},\label{eq:A weak}
\end{equation}
where $A_{s}$ and $A_{p}$ are the amplitudes of the decay process through $s$-wave and $p$-wave channels. Plugging Eq.~(\ref{eq:A weak}) into Eq.~(\ref{eq:rho^D 1/2}) and using Eqs.~(\ref{eq:dN/dOmega define}) and (\ref{eq:P_D define}), one obtains the angular distribution of $D$ in the RFP as
\begin{equation}
\frac{1}{N}\frac{dN}{d\Om^{*}}=\frac{1}{4\pi}\l(1+\al P_{P}\cos\ths\r),
\label{eq:weak_decay_angular}
\end{equation}
and its polarization vector as
\begin{equation}
\VPD=\frac{\l(\al+\VPP\cdot\hatps\r)\hatps+\be\l(\VPP\times\hatps\r)+\ga\hatps\times\l(\VPP\times\hatps\r)}{1+\al\VPP\cdot\hatps}.\label{eq:weak_decay}
\end{equation}
Here, we have introduced three decay parameters $\al$, $\be$, and $\ga$, which are defined as
\begin{equation}
\al =\frac{2\re(A_{s}^{*}A_{p})}{|A_{s}|^{2}+|A_{p}|^{2}},\
\be =\frac{2\im(A_{s}^{*}A_{p})}{|A_{s}|^{2}+|A_{p}|^{2}},\
\ga =\frac{|A_{s}|^{2}-|A_{p}|^{2}}{|A_{s}|^{2}+|A_{p}|^{2}}.\label{eq:gamma}
\end{equation}
Their values for $\Xi^-$ and $\Xi^0$ can be found from the Particle Data Group (PDG)~\cite{Tanabashi:2018oca}.

Equations (\ref{eq:weak_decay_angular}) and (\ref{eq:weak_decay}) are the well-known results for the weak decay of spin-1/2 hyperons~\cite{Lee:1957qs}. We note that, by setting $\al=\be=0$ and $\ga=\pm1$, Eqs.~(\ref{eq:weak_decay_angular}) and (\ref{eq:weak_decay}) can be reduced to Eqs.~(\ref{eq:strong1/2_angular}), (\ref{eq:strong 1/2+}), and (\ref{eq:strong 1/2-}). This is because the strong decay $1/2^{+}\to1/2^{+}0^{-}$ occurs in the pure $p$-wave mode ($\ga=-1$), whereas $1/2^{-}\to1/2^{+}0^{-}$ occurs in the pure $s$-wave mode ($\ga=1$).

\subsection{EM decay $1/2^{+}\to1/2^{+}1^{-}$}

In the EM decay $1/2^{+}\to1/2^{+}1^{-}$, the initial spin-density-matrix $\rhoiM$ is the same as Eq.~(\ref{eq:rho^i 1/2}), whereas the final spin-density-matrix $\rhofDX$ should be labeled jointly by $\la_{D}=\pm1/2$ and $\la_{X}=\pm1$. Using Eqs.~(\ref{eq:density matrix}), (\ref{eq:h matrix}), and (\ref{eq:parity conserve}), we obtain
\begin{equation}
\rhofDX=\frac{1}{8\pi}
\begin{pmatrix}
1+P_{P}\cos\ths & 0 & 0 & -P_{P}\sin\ths\\
              0 & 0 & 0 & 0\\
              0 & 0 & 0 & 0\\
 -P_{P}\sin\ths & 0 & 0 & 1-P_{P}\cos\ths
\end{pmatrix}.
\end{equation}
Here, the rows and columns are sorted in order $(\la_{D}^{(\prime)},\la_{X}^{(\prime)})=(1/2,1)$, $(-1/2,1)$, $(1/2,-1)$, and $(-1/2,-1)$. We note that only four elements of $\rho^f$ are nonzero. This is the consequence of the angular momentum conservation, which requires $|\la_D-\la_X|\leq 1/2$.

After taking the partial trace of $\rho^{f}$ over index $\la_{X}$, the spin-density matrix for $D$ is
\begin{equation}
\rhoDD=\frac{1}{8\pi}
\begin{pmatrix}
1+P_{P}\cos\ths & 0\\
              0 & 1-P_{P}\cos\ths
\end{pmatrix}.
\end{equation}
This directly leads to the angular distribution,
\begin{equation}
\frac{1}{N}\frac{dN}{d\Om^{*}}=\frac{1}{4\pi},
\end{equation}
and the polarization vector,
\begin{equation}
\VPD=-\l(\VPP\cdot\hatps\r)\hatps.\label{eq:EM_decay}
\end{equation}
These results agree with Ref.~\cite{Gatto:1958qmn}.

\subsection{Strong decay $3/2^{\pm}\to1/2^{+}0^{-}$}

For the strong decay $3/2^{\pm}\to1/2^{+}0^{-}$, the initial spin-density-matrix $\rhoiM$ is a $4\times 4$ matrix. Its diagonal elements fulfill two equations, corresponding to the normalization,
\begin{equation}
\label{eq:32one}
\rho_{\frac{3}{2}\frac{3}{2}}^{i}+\rho_{\frac{1}{2}\frac{1}{2}}^{i}+\rho_{-\frac{1}{2}-\frac{1}{2}}^{i}+\rho_{-\frac{3}{2}-\frac{3}{2}}^{i}=1,
\end{equation}
and the polarization,
\begin{equation}
\rho_{\frac{3}{2}\frac{3}{2}}^{i}+\frac{1}{3}\rho_{\frac{1}{2}\frac{1}{2}}^{i}-\frac{1}{3}\rho_{-\frac{1}{2}-\frac{1}{2}}^{i}-\rho_{-\frac{3}{2}-\frac{3}{2}}^{i}=P_{P}.\label{eq:P 3/2}
\end{equation}
Here, the polarization operator along the $z$ axis for the spin-3/2 particle is $\widehat{P}_z=\diag(1,1/3,-1/3,-1)$, and $\widehat{P}_x$ and $\widehat{P}_y$ are given by
\begin{align*}
\widehat{P}_x &= \frac{1}{3}
\begin{pmatrix}
0 & \sqrt{3} & 0 & 0 \\
\sqrt{3} & 0 & 2 & 0 \\
0 & 2 & 0 & \sqrt{3} \\
0 & 0 & \sqrt{3} & 0 \\
\end{pmatrix},\\
\widehat{P}_y &= \frac{1}{3i}
\begin{pmatrix}
0 & \sqrt{3} & 0 & 0 \\
-\sqrt{3} & 0 & 2 & 0 \\
0 & -2 & 0 & \sqrt{3} \\
0 & 0 & -\sqrt{3} & 0 \\
\end{pmatrix}.
\end{align*}
The off-diagonal elements of $\rho^i$ satisfy $\tr(\widehat{P}_x \rho^i)=\tr(\widehat{P}_y \rho^i)=0$. These two equations cannot uniquely determine all the off-diagonal elements, and Eqs.~(\ref{eq:32one}) and (\ref{eq:P 3/2}) cannot uniquely determine the four diagonal elements. To proceed, we will assume that all the off-diagonal elements are zero. This could happen if the spin degrees of freedom of all the primordial particles are thermalized so that their spin-density matrices are diagonal~\cite{Becattini:2013fla,Becattini:2016gvu}. Particularly, the diagonal elements are arranged by the thermal vorticity $\varpi$\footnote{The thermal vorticity vector is defined as $\varpi^\mu=\epsilon^{\mu\nu\rho\sigma}u_\nu\partial_\sigma (u_\rho/T)$ with $u^\mu$ as the velocity. $\varpi=|\bm\varpi|$.} and are given by~\cite{Becattini:2016gvu}
\begin{equation}
\rho_{M_{P};M_{P}}^{i} = \frac{\exp{(M_P\varpi)}}{\sum_{m=-3/2}^{3/2}\exp{(m\varpi)}}.
\end{equation}
In this case, the value of $\varpi$ and, thus, $\rho^{i}$ can be uniquely determined from $P_P$ by solving
\begin{equation}
P_P = \frac{\tanh(\varpi/2)+2\tanh(\varpi)}{3}.
\end{equation}
Furthermore, we define two parameters,
\begin{align}
\De & =\rho_{\frac{1}{2}\frac{1}{2}}^{i}+\rho_{-\frac{1}{2}-\frac{1}{2}}^{i},\\
\de & =\l(\rho_{\frac{1}{2}\frac{1}{2}}^{i}-\rho_{-\frac{1}{2}-\frac{1}{2}}^{i}\r)/\l(3P_{P}\r),
\end{align}
to characterize the initial polarization state. Both $\De$ and $\de$ are functions of $P_P$.

\begin{table*}
\caption{Daughter angular distribution and polarization vector $\VPD$ in different decay channels.}
\label{tab:Polarization vector}
\begin{ruledtabular}
\begin{tabular}{ccccc}
             & Spin and parity                                        & $(1/N)dN/d\Om^{*}$                                & $\VPD$                              & $\<\VPD\>/\VPP$ \\
\hline
Strong decay & $1/2^{+}\to1/2^{+}0^{-}$                               & $1/(4\pi)$                                        & $2\l(\VPP\cdot\hatps\r)\hatps-\VPP$ & -1/3 \\
Strong decay & $1/2^{-}\to1/2^{+}0^{-}$                               & $1/(4\pi)$                                        & $\VPP$                              & 1 \\
Strong decay & $3/2^{+}\to1/2^{+}0^{-}$                               & $3\l[1-2\De/3-\l(1-2\De\r)\cos^{2}\ths\r]/(8\pi)$ & Eq.~(\ref{eq:strong 3/2+})          & 1 \\
Strong decay & $3/2^{-}\to1/2^{+}0^{-}$                               & $3\l[1-2\De/3-\l(1-2\De\r)\cos^{2}\ths\r]/(8\pi)$ & Eq.~(\ref{eq:strong 3/2-})          & -3/5 \\
Weak decay   & $1/2^{\phantom{+}}\to1/2^{\phantom{+}}0^{\phantom{+}}$ & $\l(1+\al P_{P}\cos\ths\r)/(4\pi)$                & Eq.~(\ref{eq:weak_decay})           & $(2\ga+1)/3$ \\
EM decay     & $1/2^{+}\to1/2^{+}1^{-}$                               & $1/(4\pi)$                                        & $-\l(\VPP\cdot\hatps\r)\hatps$      & -1/3 \\
\end{tabular}
\end{ruledtabular}
\end{table*}

After some calculations similar to the above subsections, one can obtain the angular distribution of $D$,
\begin{equation}
\frac{1}{N}\frac{dN}{d\Om^{*}}=\frac{3}{8\pi}\l[1-\frac{2}{3}\De-\l(1-2\De\r)\cos^{2}\ths\r].
\end{equation}
This result is analogous to the spin alignment of the vector meson~\cite{Liang:2004xn,Abelev:2008ag,Zhou:2019lun,Singh:2018uad}. If the parent is polarized, we expect $\Delta<1/2$, and the daughter's angular distribution becomes anisotropic. The polarization vector of $D$ is obtained to be
\begin{widetext}
\begin{equation}
\VPD=\frac{-4\de\l(\VPP\cdot\hatps\r)\hatps+\l[1-2\de-(1-10\de)\l(\hatPP\cdot\hatps\r)^{2}\r]\VPP}{1-2\De/3-\l(1-2\De\r)\l(\hatPP\cdot\hatps\r)^{2}},\label{eq:strong 3/2+}
\end{equation}
and
\begin{equation}
\VPD=\frac{2\l[1-4\de-(1-10\de)\l(\hatPP\cdot\hatps\r)^{2}\r]\l(\VPP\cdot\hatps\r)\hatps-\l[1-2\de-(1-10\de)\l(\hatPP\cdot\hatps\r)^{2}\r]\VPP}{1-2\De/3-\l(1-2\De\r)\l(\hatPP\cdot\hatps\r)^{2}}.\label{eq:strong 3/2-}
\end{equation}
\end{widetext}
Here, Eqs.~(\ref{eq:strong 3/2+}) and (\ref{eq:strong 3/2-}) are for $3/2^{+}\to1/2^{+}0^{-}$ and $3/2^{-}\to1/2^{+}0^{-}$, respectively. We note here that, if the initial polarization is small ($P_{P}\simeq 0$), we have
\begin{align*}
\rho_{M_{P};M_{P}}^{i} &=\frac{1}{4}+\frac{1}{4}M_{P}\varpi +O(\varpi^2),\nonumber\\
P_P &=\frac{5}{6}\varpi +O(\varpi^2),\nonumber
\end{align*}
and, thus, $\Delta=1/2$ and $\delta=1/10$. In this case, Eqs.~(\ref{eq:strong 3/2+}) and (\ref{eq:strong 3/2-}) are dramatically simplified
\begin{equation}
\VPD=\frac{6}{5}\left[\VPP-\frac{1}{2}\l(\VPP\cdot\hatps\r)\hatps\right],
\end{equation}
and
\begin{equation}
\VPD=-\frac{6}{5}\left[\VPP-\frac{3}{2}\l(\VPP\cdot\hatps\r)\hatps\right],
\end{equation}
respectively. On the other hand, if the parent is ultrapolarized ($|\VPP|\simeq 1$), the initial density-matrix $\rho^{i}$ is condensed at the $M_{P}=3/2$ or the $-3/2$ state, then, we have $\De=\de=0$, and Eqs.~(\ref{eq:strong 3/2+}) and (\ref{eq:strong 3/2-}) are reduced to
\begin{equation}
\VPD=\VPP,
\end{equation}
and
\begin{equation}
\VPD=2\l(\VPP\cdot\hatps\r)\hatps-\VPP,
\end{equation}
respectively. In our simulations presented in the next section, the initial polarization can take an arbitrary value, thus, we use Eqs.~(\ref{eq:strong 3/2+}) and (\ref{eq:strong 3/2-}).

Before we end this section, we summarize the above results in Table~\ref{tab:Polarization vector}. The last column shows the factor $C=\langle\VPD\rangle/\VPP$ in Eq.~(\ref{eq:linear rule}) where the averaged polarization $\langle\VPD\rangle$ is obtained by
\begin{equation}
\langle\VPD\rangle=\int d\Omega^* \frac{1}{N}\frac{dN}{d\Om^{*}} \VPD.
\end{equation}
The results for $\langle\VPD\rangle$ are consistent with Ref.~\cite{Becattini:2016gvu}.

\section{Numerical simulation} \label{sec3}

In this section, we present Monte Carlo simulations to study the effect of feed-downs on $\La$ polarization using the formalism obtained in the last section. All the simulations are aimed at noncentral Au + Au collisions at $\sNN=200$ GeV in 10--60\% centrality.

We first describe the method for determining the yields and the kinetic distributions of the primordial particles in Sec.~\ref{sec3A}. Then, a series of simulation results are presented in Sec.~\ref{sec3B}. Some discussions are given in Sec.~\ref{sec3C}.

\subsection{Simulation setup} \label{sec3A}

In this subsection, we set up the input information for our simulation, including the yields and the kinetic distributions of the primordial particles. The particle species that we include in our simulations are listed in Table~\ref{tab:particle}. Their primordial yields are determined by the statistical thermodynamic model, presented as a ratio to the yield of $\La$. We employ the grand-canonical ensemble in which the partition function is given by
\begin{equation}
\ln Z=\sum_{\text{species }i}\frac{g_{i}V}{\l(2\pi\r)^{3}}\int d^{3}p\ln\l[1\pm\exp\l(\frac{E_{i}-\mu_{i}}{\Tch}\r)\r]^{\pm1},\label{eq:ZGC}
\end{equation}
where $g_{i}$ is the spin degeneracy factor, $V$ is the volume of the thermal system, $\Tch$ is the chemical freeze-out temperature, $E_i=\sqrt{p^2+m_i^2}$ is the particle energy, and $\mu_{i}=B_{i}\mu_{B}+S_{i}\mu_{S}+Q_{i}\mu_{Q}$ is the chemical potential in which $B_{i}$, $S_{i}$, and $Q_{i}$ are the baryon, strangeness, and charge numbers of particle species $i$, and $\mu_{B}$, $\mu_{S}$, and $\mu_{Q}$ are the corresponding chemical potentials. The plus and minus signs correspond to fermions and bosons, respectively. From Eq.~(\ref{eq:ZGC}), the primordial yield number of particle species $i$ can be calculated by
\begin{equation}
N_{i}=\Tch\frac{\pd(\ln Z)}{\pd\mu_{i}}.\label{eq:NGC}
\end{equation}
We adopt the \texttt{THERMUS} package~\cite{Wheaton:2004qb} to calculate Eqs.~(\ref{eq:ZGC}) and (\ref{eq:NGC}) with the calculational parameters taking the same values as in Ref.~\cite{Adamczyk:2017iwn} where the yields of $\pi^{\pm}$, $K^{\pm}$, $p$, $\bar{p}$, $\La$, $\bar{\La}$, $\Xi^{-}$, and $\bar{\Xi}^{+}$ are fitted to the experimental data. We, thus, obtain the multiplicities of the primordial particles for the 10--60\% central Au + Au collision at $\sNN=200$ GeV. The results are listed in the second column of Table~\ref{tab:particle} as the ratios to the primordial $\La$ yield.

In Table~\ref{tab:particle}, we also list particles' spin, parity, and main decay channels that can contribute to final $\La$ hyperons. Using the $N_{i}/N_{\La}$ data and the branch ratio for each decay channel in the PDG~\cite{Tanabashi:2018oca}, we find that only around 21\% of the final $\La$s are primordial, and the others are produced by decays from high-lying strange baryons: about 15\% from the decay of primordial $\Si^{0}$, 30\% from primordial $\Si^{*0}$, $\Si^{*+}$, and $\Si^{*-}$, 14\% from primordial $\Xi^{0}$ and $\Xi^{-}$, 10\% from primordial $\Xi^{*0}$ and $\Xi^{*-}$, and 10\% from other higher-lying resonance states.

\begin{table}[t]
\caption{The primordial yield ratio $N_{i}/N_{\La}$, spin, parity, and decay channels of strange particles.}
\label{tab:particle}
\begin{ruledtabular}
\begin{tabular}{cccl}
            & $N_{i}/N_{\La}$ & Spin and parity & Decay channel            \\
\hline
$\La$       & 1               & $1/2^{+}$       & -                        \\
$\La(1405)$ & 0.236           & $1/2^{-}$       & $\Si^{0}\pi$             \\
$\La(1520)$ & 0.265           & $3/2^{-}$       & $\Si^{0}\pi$             \\
$\La(1600)$ & 0.098           & $1/2^{+}$       & $\Si^{0}\pi$             \\
$\La(1670)$ & 0.061           & $1/2^{-}$       & $\Si^{0}\pi$, $\La\eta$  \\
$\La(1690)$ & 0.112           & $3/2^{-}$       & $\Si^{0}\pi$             \\
$\Si^{0}$   & 0.686           & $1/2^{+}$       & $\La\ga$                 \\
$\Si^{*0}$  & 0.533           & $3/2^{+}$       & $\La\pi$                 \\
$\Si^{*+}$  & 0.535           & $3/2^{+}$       & $\La\pi$, $\Si^{0}\pi$   \\
$\Si^{*-}$  & 0.524           & $3/2^{+}$       & $\La\pi$, $\Si^{0}\pi$   \\
$\Si(1660)$ & 0.068           & $1/2^{+}$       & $\La\pi$, $\Si^{0}\pi$   \\
$\Si(1670)$ & 0.125           & $3/2^{-}$       & $\La\pi$, $\Si^{0}\pi$   \\
$\Xi^{0}$   & 0.343           & $1/2^{+}$       & $\La\pi$                 \\
$\Xi^{-}$   & 0.332           & $1/2^{+}$       & $\La\pi$                 \\
$\Xi^{*0}$  & 0.228           & $3/2^{+}$       & $\Xi\pi$                 \\
$\Xi^{*-}$  & 0.224           & $3/2^{+}$       & $\Xi\pi$                 \\
\end{tabular}
\end{ruledtabular}
\end{table}

\begin{figure}
\centering \includegraphics[width=1\columnwidth]{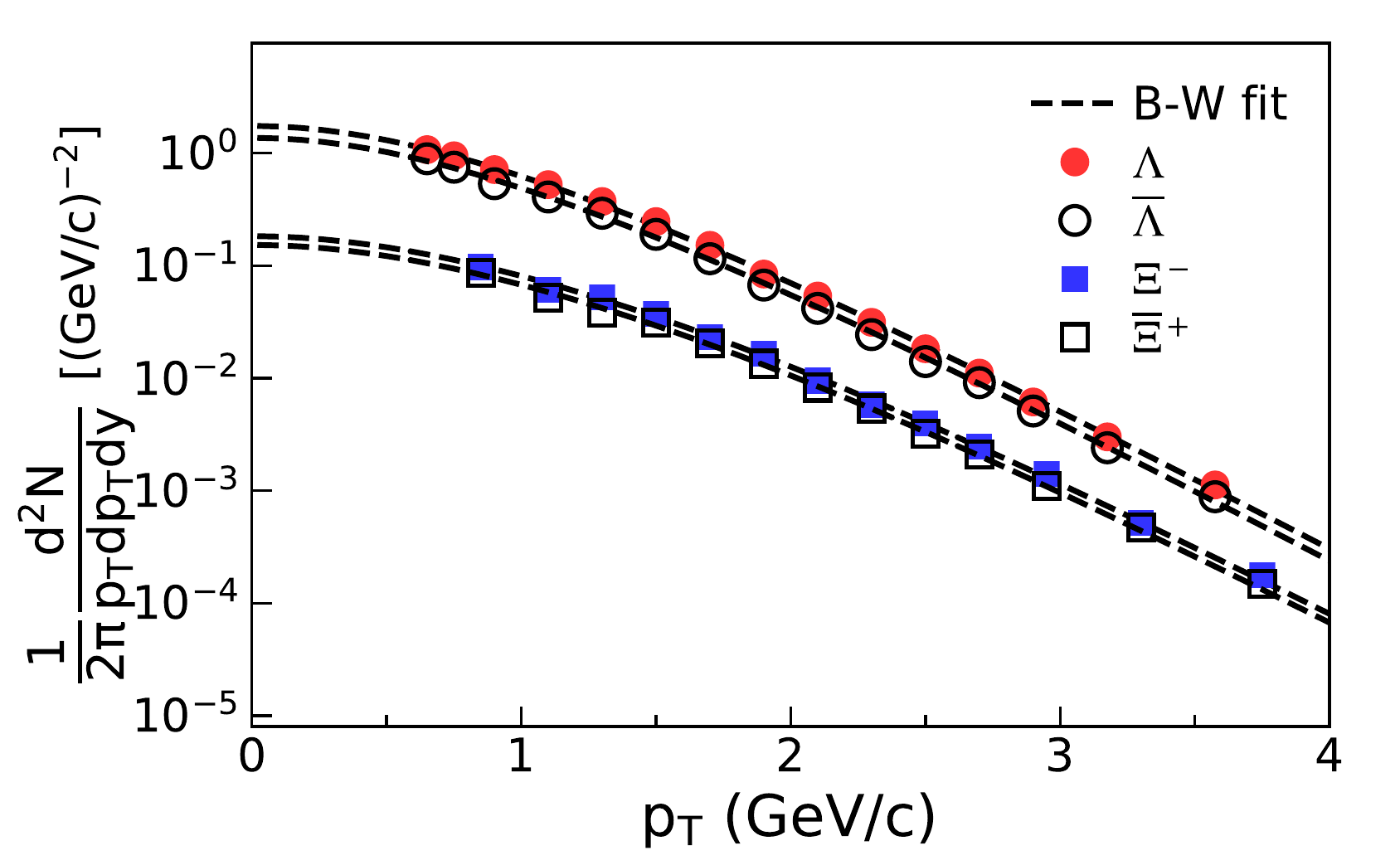}
\caption{Blast-wave fits (dashed curves) to the data of the STAR Collaboration~\cite{Adams:2006ke} for $\pT$ spectra of $\La$, $\bar{\La}$, $\Xi^{-}$, and $\bar{\Xi}^{+}$ in 10--60\% central Au + Au collisions at $\sNN=200$ GeV.}
\label{fig:pt-fit}
\end{figure}

As for the momentum distributions of the primordial particles, we assume the rapidity distribution $dN/d\eta$ to be flat at 200 GeV, and the transverse momentum $\pT$ is generated by the blast-wave model~\cite{Schnedermann:1993ws} using the following equation:
\begin{equation}
\frac{d^{2}N}{\pT d\pT dy}\propto\int_{0}^{1}\tilde{r}d\tilde{r}\mT I_{0}\l(\frac{\pT\sinh\rho}{\Tkin}\r)K_{1}\l(\frac{\mT\cosh\rho}{\Tkin}\r),\label{eq:pt_spectra}
\end{equation}
where $\mT=\sqrt{\pT^{2}+m_{0}^{2}}$ is the transverse mass and $m_{0}$ is the particle mass. For resonance particles, $m_{0}$ is sampled according to the Breit-Wigner distribution with the central mass and width from PDG~\cite{Tanabashi:2018oca}. $\Tkin$ is the kinetic freeze-out temperature, $I_{0}$ and $K_{1}$ are the modified Bessel functions, and $\rho$ is the transverse rapidity parametrized as
\begin{equation}\label{eq:radialrapidity}
\tanh\rho=\frac{2+l}{2}\<\be\>\tilde{r}^{l},
\end{equation}
where $\<\beta\>$ is the average radial flow in the thermal area, $\tilde{r}=r/r_{\mathrm{max}}$ is the reduced radius, and $l$ is the exponent of the flow profile. In our simulation, the blast-wave parameters take the values of $\Tkin=140$ MeV, $\<\be\>=0.445$, and $l=1.21$, which are determined by fitting the experimental data~\cite{Adams:2006ke} for $\pT$ spectra of $\La$, $\bar{\La}$, $\Xi^{-}$, and $\bar{\Xi}^{+}$ in 10--60\% central Au + Au collisions at $\sNN=200$ GeV. The fitting result is shown in Fig.~\ref{fig:pt-fit}.

\begin{figure}
\centering \includegraphics[width=1\columnwidth]{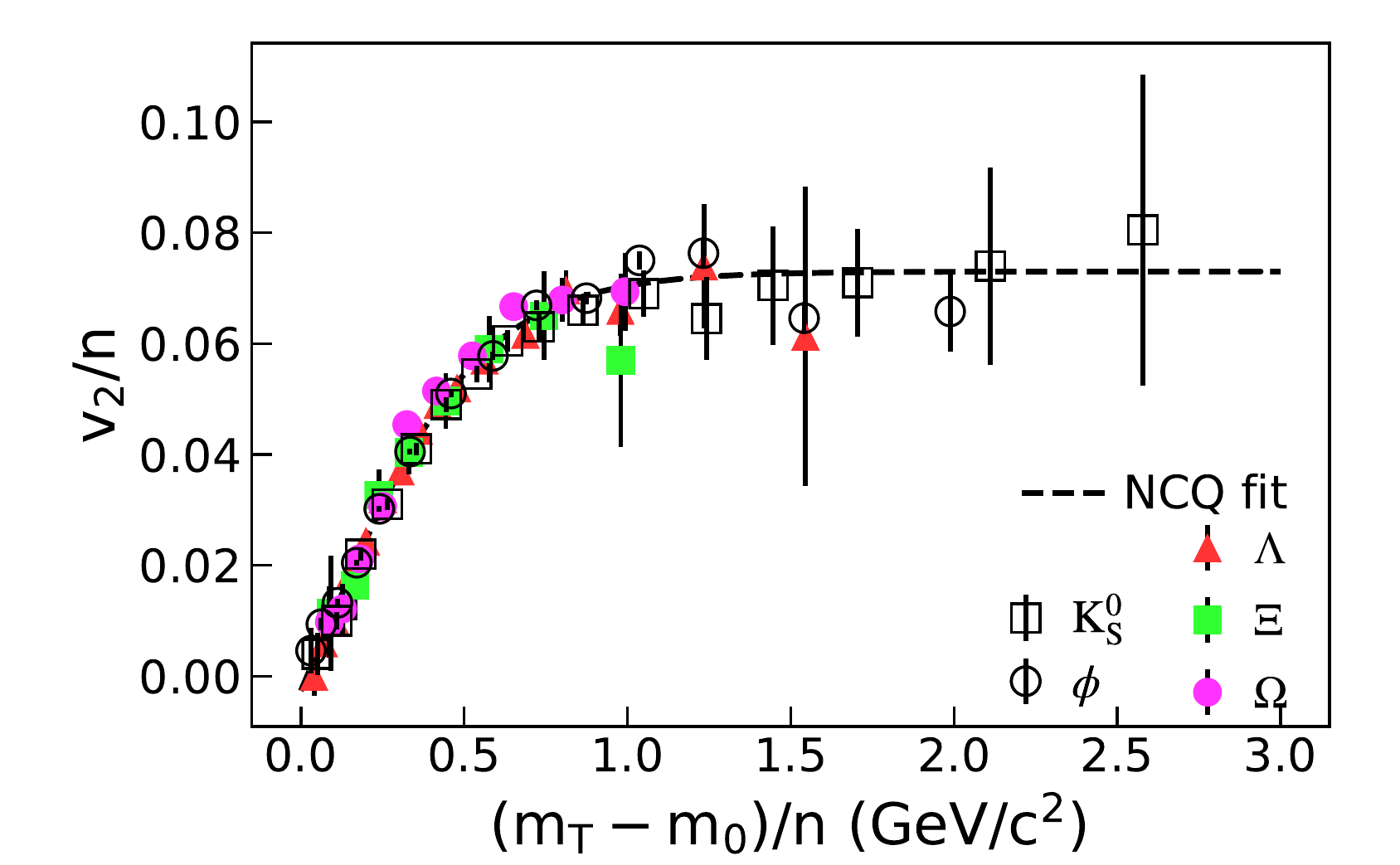}
\caption{NCQ fit (dashed curve) to the data of the STAR Collaboration~\cite{Adams:2003am,Adams:2005zg,Adamczyk:2015ukd} for elliptic flows of $K_{S}^{0}$, $\phi$, $\La$, $\Xi$, and $\Om$ in the minimum-bias Au + Au collisions at $\sNN=200$ GeV.}
\label{fig:v2-fit}
\end{figure}

When sampling the azimuthal angle, all primordial particles are allowed to have their elliptic flows. The value of $v_{2}$ is calculated  by the following function inspired by the number of constituent quark (NCQ) scaling law~\cite{Dong:2004ve}:
\begin{equation}
v_{2}/n=\frac{a}{1+\exp\l\{-\l[\l(\mT-m_{0}\r)/n-b\r]/c\r\}}-d,
\end{equation}
where $n$ is the number of constituent quarks in a hadron. The parameter values are $a=0.133$, $b=0.066$ GeV/$c$, $c=0.238$ GeV/$c$, and $d=0.06$, which are determined by fitting the elliptic flows of $K_{S}^{0}$, $\phi$, $\La$, $\Xi$, and $\Om$ in the minimum-bias Au + Au collisions at $\sNN=200$ GeV~\cite{Adams:2003am,Adams:2005zg,Adamczyk:2015ukd} as shown in Fig.~\ref{fig:v2-fit}.

We note that, in the above fittings, the experimental data for the $\pT$ spectra and the elliptic flows include the contribution from feed-down decays; only the $\pT$ spectra of $\La$($\bar{\La}$) are corrected by excluding the weak decays from $\Xi$ and $\Om$~\cite{Adams:2006ke}. Besides, the available $v_{2}$ data for (multi)strangeness particles are from minimum-bias (0--80\%) events, whereas our simulation is for 10--60\% central collisions. However, as we checked by varying the yields, $\pT$ spectra, and $v_{2}$ data, these mismatches have only a minor impact on our results presented in the following subsections. Nevertheless, if more data from the experiments are released, our fitting can be gradually improved.

\subsection{Feed-down effect on $\La$ polarization} \label{sec3B}

In this subsection, we perform a series of Monte Carlo simulations to study the effect of particle decay on $\La$ polarization. For each plot shown in this section, $10^{9}$ primordial particles are sampled. Their species and momenta are determined randomly by the yield ratio $N_{i}/N_{\La}$ and the momentum distribution in last subsection. The polarizations of primordial particles are input as function of the particle's azimuthal angle. All spin-1/2 primordial particles are assumed to have the same polarization with that of the primordial $\La$'s, whereas, for spin-3/2 primordial particles, their polarization vector $P_{3/2}$ is determined from the spin-1/2 polarization $P_{1/2}$ by solving
\begin{align}
P_{3/2} & = [\tanh(\varpi/2) + 2\tanh(\varpi)] / 3, \label{eq:prim3/2}\\
P_{1/2} & = \tanh(\varpi/2). \label{eq:prim1/2}
\end{align}
These relations are obtained by assuming thermal equilibrium for the spin degree of freedom so that the polarization is determined by the thermal-vorticity $\varpi$~\cite{Becattini:2016gvu}. The angular distribution and polarization of the daughter particle in each decay channel are calculated by equations obtained in Sec.~\ref{sec2}, and the daughter's momentum is boosted to the laboratory frame by the parent's momentum. In all the simulations, primordial particles are generated in ranges of $|\eta|<2$ and $0<\pT<8$ GeV/$c$, whereas only the final $\La$s in ranges of $|\eta|<1$ and $0.5<\pT<4$ GeV/$c$ are selected for analysis. In the present paper, we do not distinguish $\La$ and $\bar\La$.

\begin{figure}[t]
\centering \includegraphics[width=1\columnwidth]{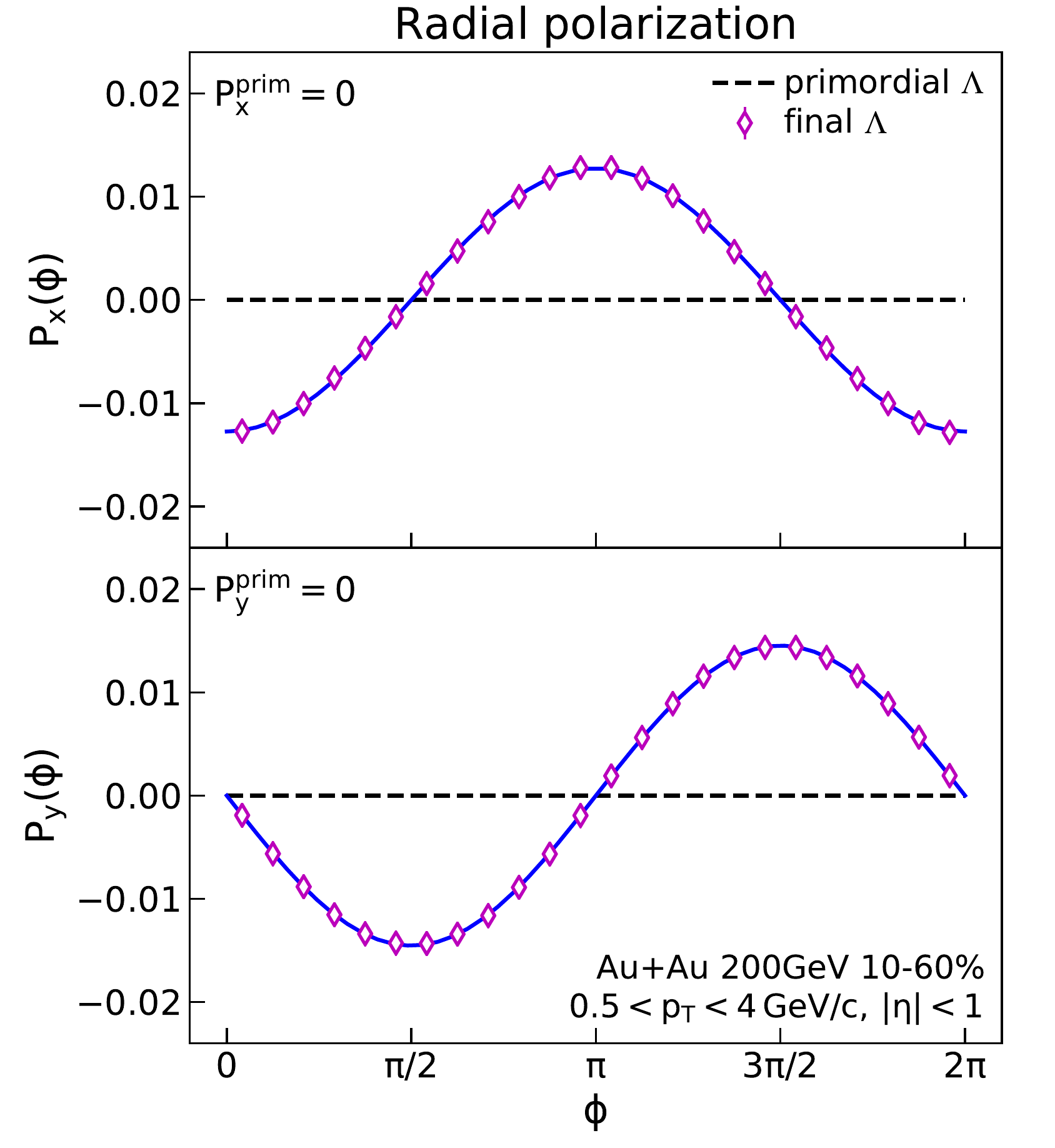}
\caption{Radial polarization of final $\La$'s (diamond points) generated from feed-downs with zero input of primordial polarization (black dashed curves). The blue curves represent the harmonic fit to the final $\La$ polarization. The simulation is for Au + Au collisions at $\sNN=200$ GeV.}
\label{fig:decay0}
\end{figure}

First of all, let us study the case that the input primordial polarization is zero $\mathbf{P}^{\mathrm{prim}}=(0,0,0)$. The final $\La$ polarization is presented as a function of $\La$'s azimuthal angle in Fig.~\ref{fig:decay0}. We find the final $\La$s are polarized, even though the polarizations of all primordial particles are zero. This is due to the weak decay of $\Xi^{-}$ and $\Xi^{0}$. From Eq.~(\ref{eq:weak_decay}), one can see that $\La$s decaying from unpolarized $\Xi$ could have a radial polarization $\mathbf{P}_{\La}=\al_{\Xi}\hatps$, where $\hatps$ is $\La$'s momentum direction in the rest frame of $\Xi$. After boosting back to the laboratory frame, this radial polarization gives Fig.~\ref{fig:decay0}. The longitudinal component $P_{z}$ is zero averaged in the symmetric rapidity range of $-1<\eta<1$ and is not shown in the figure. The blue curves in Fig.~\ref{fig:decay0} show the fits to final $\La$ polarization by the following equations:
\begin{align}
P_{x}^{\mathrm{final}}(\phi) & =K_{1x}\cos\phi, \label{eq:P_x_primary-1}\\
P_{y}^{\mathrm{final}}(\phi) & =K_{1y}\sin\phi, \label{eq:P_y_primary-1}
\end{align}
where $\phi$ is the azimuthal angle of $\La$'s momentum with respect to the $x$ axis and $(\cos\phi, \sin\phi)$ is the unit vector in the radial direction. The coefficient values are $K_{1x}=-0.127$ and $K_{1y}=-0.145$, which mean that the final $\La$s are polarized pointing almost opposite to $\La$'s momentum direction. This is consistent with the fact that $\al_{\Xi^-}$ and $\al_{\Xi^0}$ are negative~\cite{Tanabashi:2018oca}. The difference between $K_{1x}$ and $K_{1y}$ is due to the existence of the parent's elliptic flow. We have checked that, if we remove $\Xi^{-}$, $\Xi^{0}$, $\Xi^{*-}$, and $\Xi^{*0}$ from the primordial particle species in the simulation, all components of final $\La$ polarization are zero; and if we set the primordial $v_2$ to be zero, the values of $K_{1x}$ and $K_{1y}$ are equal.

Next, we input nonzero primordial polarization. Three issues are taken into consideration. They are as follows: (1) the in-plane to out-of-plane differences of the polarization at midrapidity, (2) the transverse local polarization at positive or negative rapidity, and (3) the longitudinal local polarization at midrapidity. The input primordial polarization of $\La$ is parametrized as a superposition of harmonic functions of $\La$'s azimuthal-angle $\phi$ in the following form:
\begin{alignat}{3}
P_{x}^{\mathrm{prim}} & =        & f_{1x}\sin\phi & ,                      \label{eq:P_x_primary}\\
P_{y}^{\mathrm{prim}} & = f_{0}- & f_{1y}\cos\phi & + && f_{2}\cos(2\phi), \label{eq:P_y_primary}\\
P_{z}^{\mathrm{prim}} & =        &                &   && f_{z}\sin(2\phi). \label{eq:P_z_primary}
\end{alignat}
Other particles in the simulation are given polarizations as discussed around Eqs.~(\ref{eq:prim3/2}) and (\ref{eq:prim1/2}). In Eqs.~(\ref{eq:P_x_primary})--(\ref{eq:P_z_primary}), $f_0$ stands for the global polarization, the term $f_{2}\cos(2\phi)$ characterizes a difference in $P_{y}^{\mathrm{prim}}$ at midrapidity from in-plane ($\phi=0$ or $\pi$) to out-of-plane ($\phi=\pi/2$ or $3\pi/2$) directions; $(f_{1x}\sin\phi,-f_{1y}\cos\phi)$ is the transverse local polarization, and $f_{z}\sin(2\phi)$ is the longitudinal local polarization. In the following simulation, the coefficient values are taken to be $f_{0}=f_{2}=0.0025$ and $f_{z}=0.002$, for instance, which are of the typical magnitude close to the current preliminary experimental data~\cite{Niida:2018hfw}. The coefficients $f_{1x}$ and $f_{1y}$ are assumed linear relations with rapidity  $f_{1x}=f_{1y}=2\overline{f}_1\eta$, where $\overline{f}_1$ is the mean value of $|f_{1x}|$ and $|f_{1y}|$ in rapidity region $|\eta|<1$. Its value takes $\overline{f}_1=0.1$, estimated in Ref.~\cite{Xia:2018tes}. It is important to point out that the coefficients $f_{1x}$ and $f_{1y}$ are found to be rapidity odd, whereas $f_{0}$, $f_{2}$, and $f_{z}$ are rapidity even, see the discussions in Refs.~\cite{Xia:2018tes,Wei:2018zfb}. Because of their different dependences on rapidity, when $\La$ polarization is averaged on symmetric rapidity region $-1<\eta<1$, terms related to $f_{1x}$ and $f_{1y}$ get canceled, and the effects of $f_{0}$, $f_{2}$, and $f_{z}$ survive. In contract, to extract the terms with $f_{1x}$ and $f_{1y}$, $\La$ polarization are averaged separately in the regions of $-1<\eta<0$ and $0<\eta<1$ or, equivalently, averaged with the weight of the sign of rapidity.

\begin{figure}[t]
\centering \includegraphics[width=1\columnwidth]{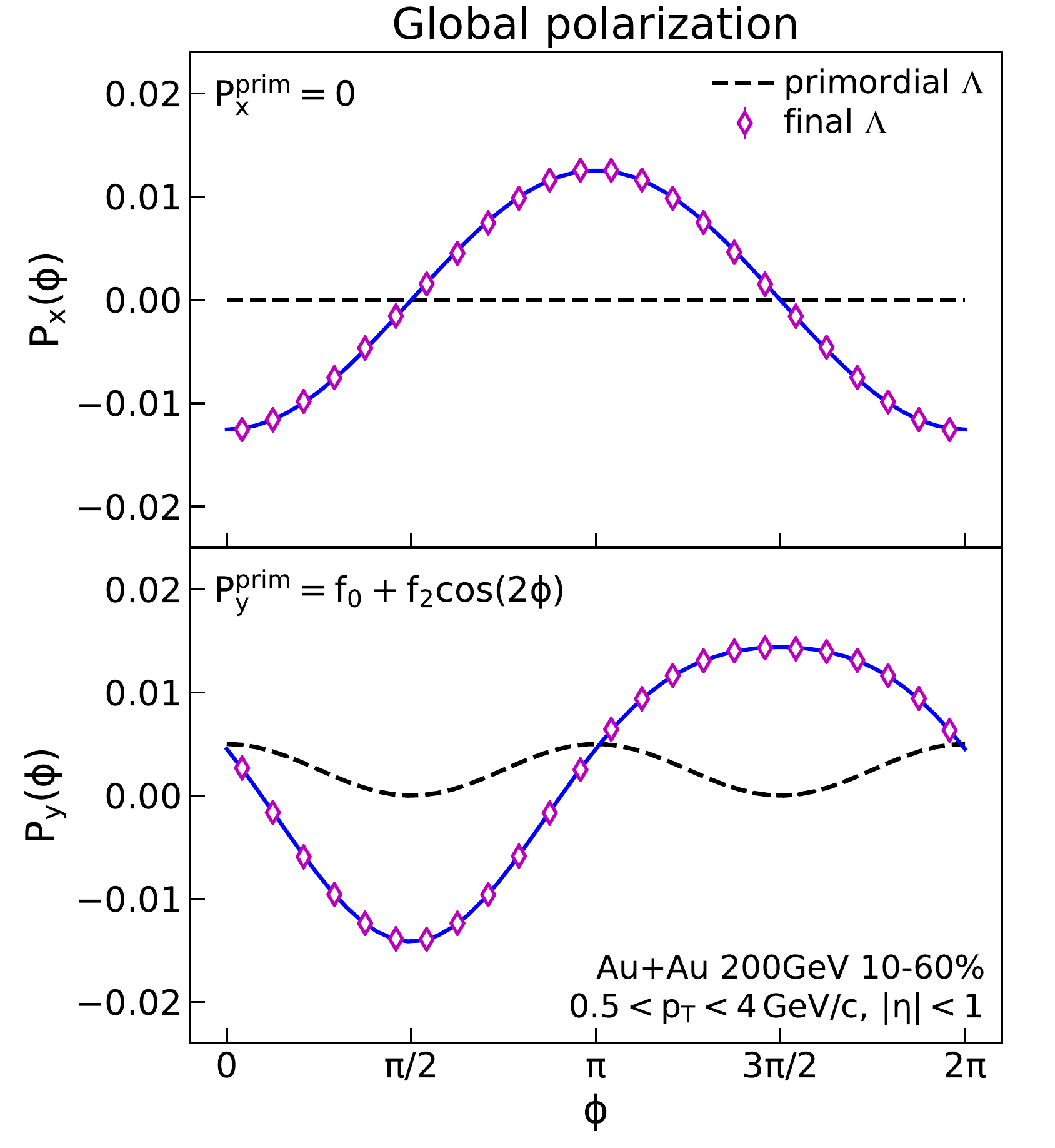}
\caption{Effect of the feed-downs on the global $\La$ polarization with in-plane and out-of-plane differences. Compared to the primordial $\La$ polarization (black dashed curves), the final $\La$ polarization after decay (diamond points) and the harmonic fit to the final $\La$ polarization (blue curves) are shown. The simulation is for Au + Au collisions at $\sNN=200$ GeV.}
\label{fig:decayF0}
\end{figure}

\begin{figure}[t]
\centering \includegraphics[width=1\columnwidth]{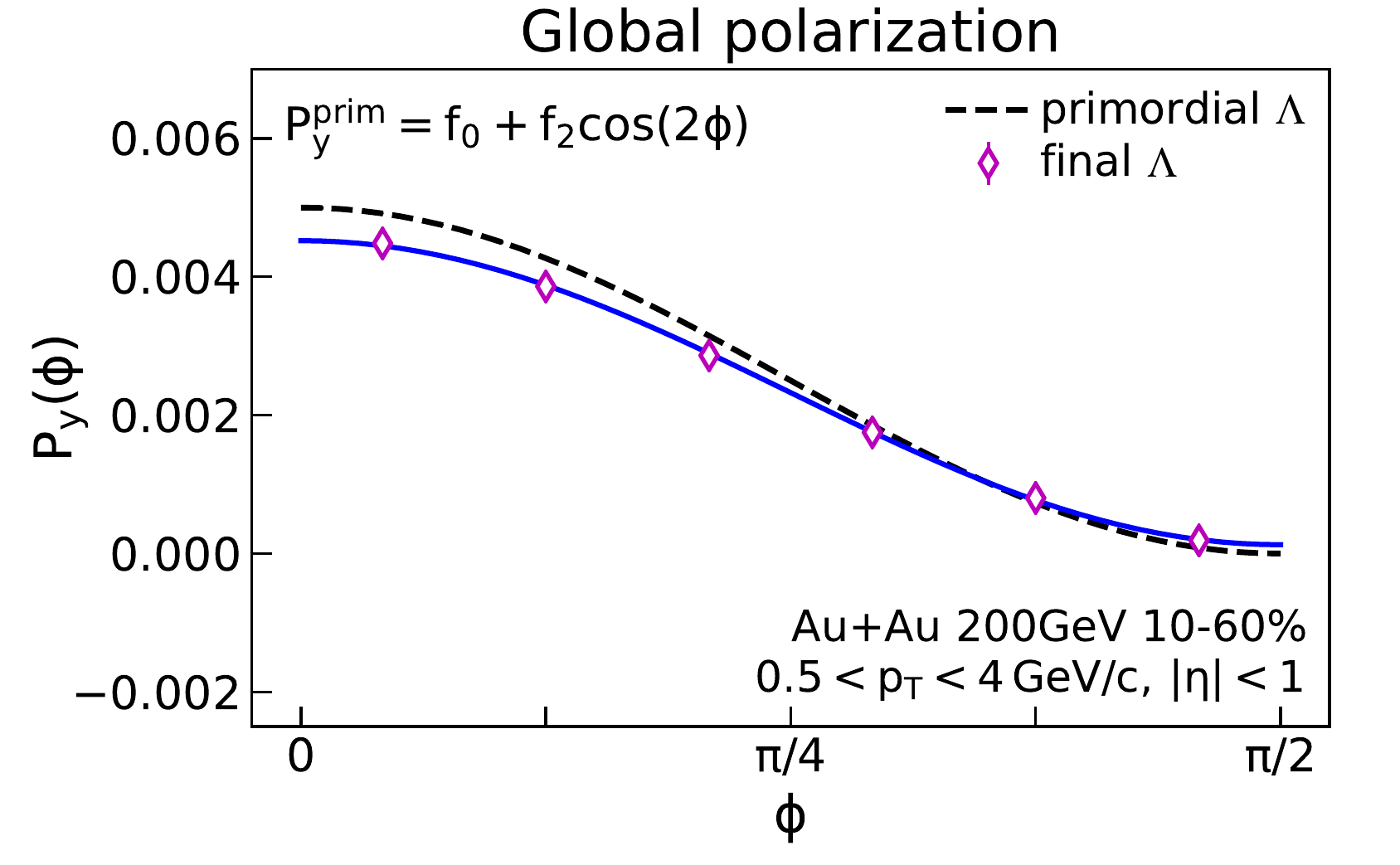}
\caption{The same as Fig.~\ref{fig:decayF0} with $\La$'s azimuthal angle folded into range $(0,\pi/2)$.}
\label{fig:decayF0fold}
\end{figure}

\begin{figure}[t]
\centering \includegraphics[width=1\columnwidth]{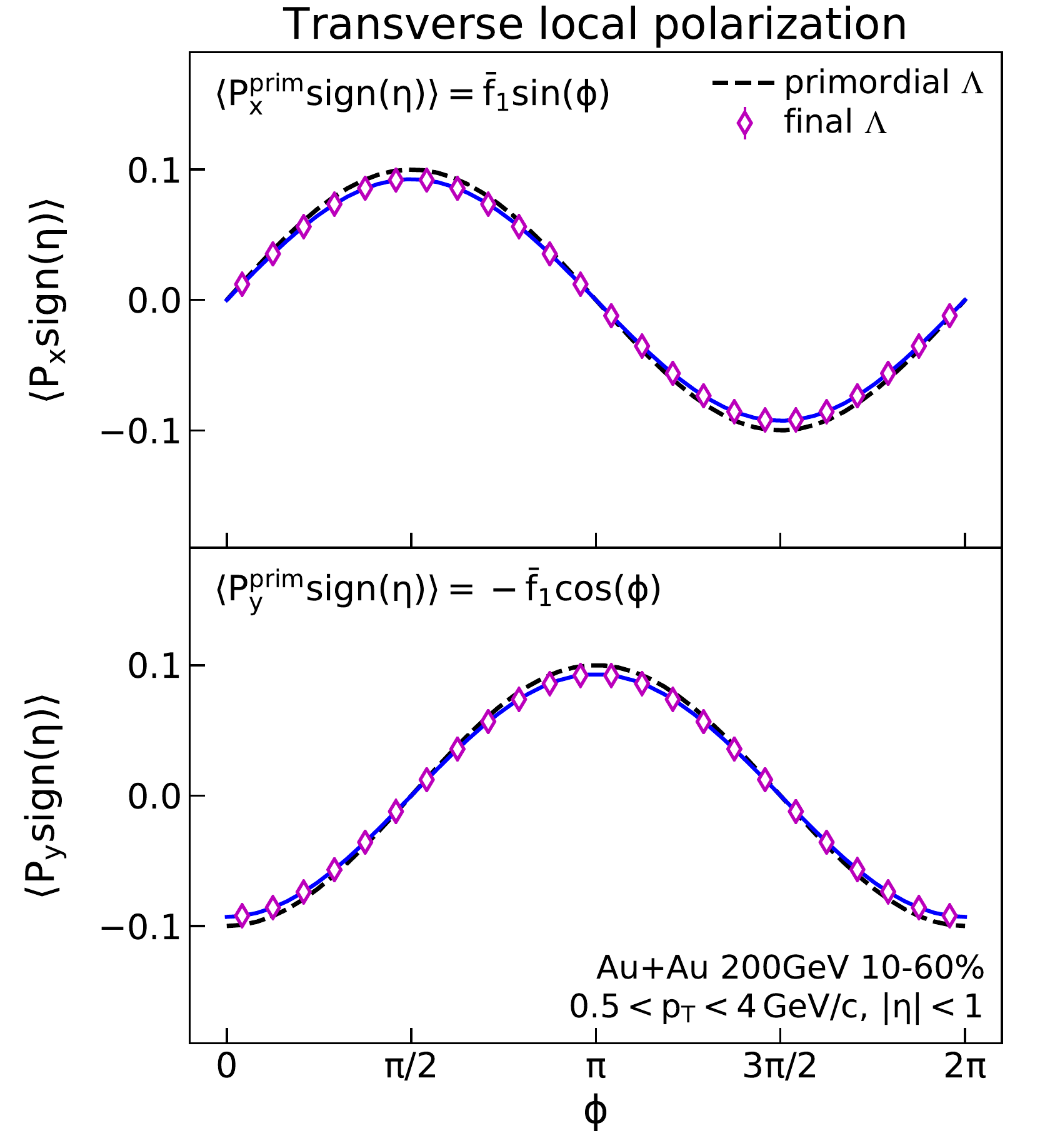}
\caption{Transverse local $\La$ polarization: The primordial $\La$ polarization (black dashed curves), the final $\La$ polarization after decay (diamond points), and the harmonic fit to the final $\La$ polarization (blue curves). The simulation is for Au + Au collisions at $\sNN=200$ GeV.}
\label{fig:decayF1}
\end{figure}

\begin{figure}[t]
\centering \includegraphics[width=1\columnwidth]{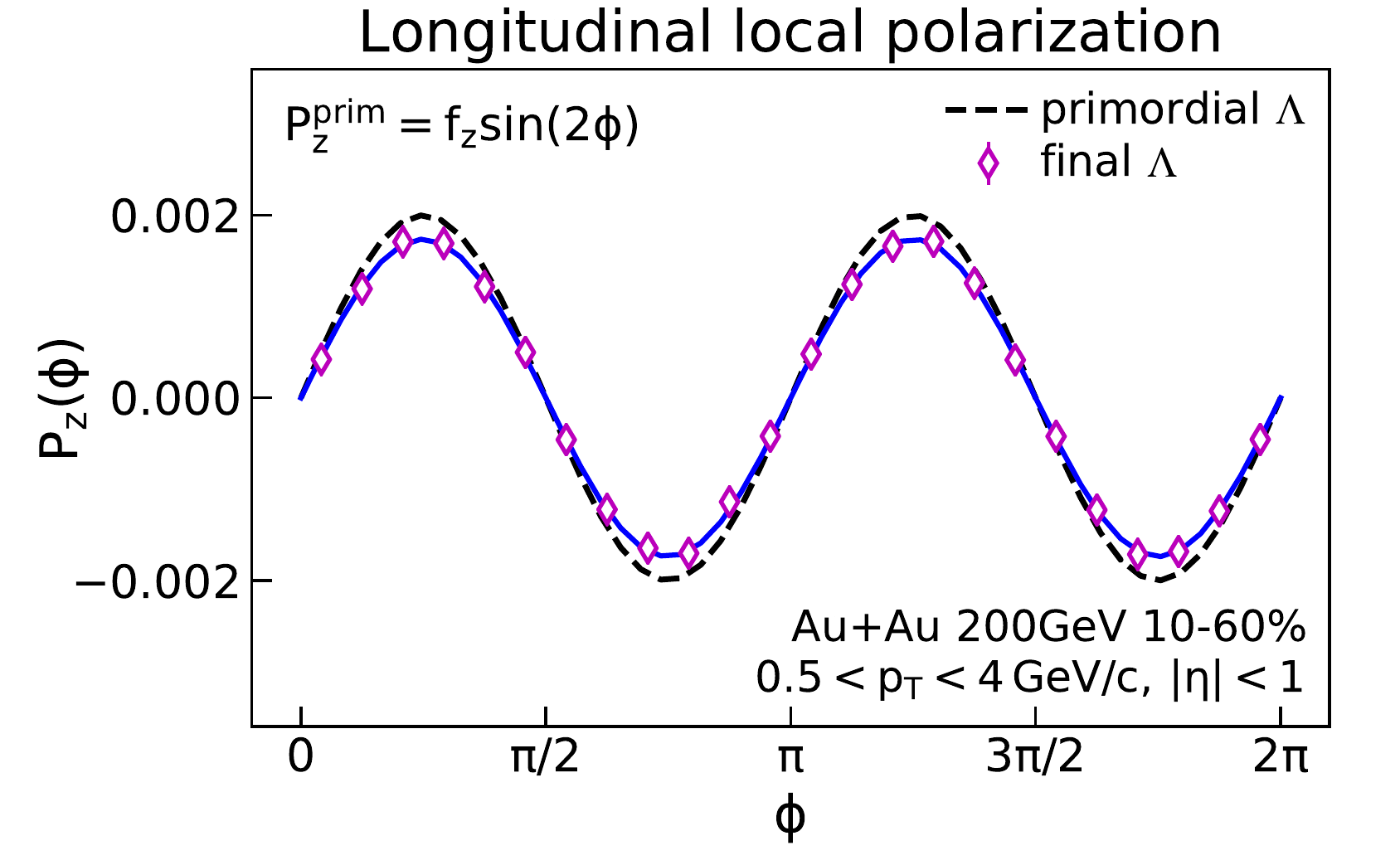}
\caption{Longitudinal local $\La$ polarization: The primordial $\La$ polarization (black dashed curve), the final $\La$ polarization after decay (diamond points), and the harmonic fit to the final $\La$ polarization (blue curve). The simulation is for Au + Au collisions at $\sNN=200$ GeV.}
\label{fig:decayFz}
\end{figure}

Figure~\ref{fig:decayF0} shows the effect of the feed-downs on the global $\La$ polarization with a difference from in-plane to out-of-plane directions. We find that the output signal for the final $\La$ polarization is the superposition of the radial polarization and the global polarization which can be well fitted by
\begin{alignat}{2}
P_{x}^{\mathrm{final}}(\phi) & =       &&K_{1x}\cos\phi,\label{eq:P_x_primary-1}\\
P_{y}^{\mathrm{final}}(\phi) & = F_{0}+&&K_{1y}\sin\phi+F_{2}\cos(2\phi).\label{eq:P_y_primary-1}
\end{alignat}
We can see that the peak magnitude of $K_{1y}\sin\phi$ is much larger than that of $F_{0}+F_{2}\cos(2\phi)$, so the final polarization follows roughly a sine shape.

In the experimental measurement for the in-plane to out-of-plane differences of the global polarization~\cite{Niida:2018hfw}, the azimuthal angle of $\La$ is defined in range of (0, $\pi/2$) by folding twice the entire range of (0, $2\pi$). After applying the same analysis as the experimental measurement, we obtain the result shown in Fig.~\ref{fig:decayF0fold} in which the radial polarization $K_{1y}\sin\phi$ vanishes, and we can see that the final $\La$ polarization is suppressed compared with the primordial one. The blue curve in  Fig.~\ref{fig:decayF0fold} is the fit to the final $\La$ polarization by
\begin{equation}
P_{y}^{\mathrm{final}}(\phi) = F_{0}+F_{2}\cos(2\phi),
\end{equation}
We find the suppression factors for the isotropic polarization and the in-plane to out-of-plane differences are $F_{0}/f_0=0.93$ and $F_2/f_2=0.88$, respectively.

Figure~\ref{fig:decayF1} shows the effect of the feed-downs on the transverse local $\La$ polarization where the data are averaged separately in range of $\eta>0$ and $\eta<0$ and then combined with the weight of the sign of $\eta$, namely,
\begin{align}
\< P_{x} \mathrm{sign}(\eta) \> & \equiv \frac{P_{x}(\eta>0)-P_{x}(\eta<0)}{2}, \\
\< P_{y} \mathrm{sign}(\eta) \> & \equiv \frac{P_{y}(\eta>0)-P_{y}(\eta<0)}{2}.
\end{align}
The blue curves in Fig.~\ref{fig:decayF1} are the fit to the final $\La$ polarization by the following equations:
\begin{align}
\< P_{x}^{\mathrm{final}} \mathrm{sign}(\eta) \> & = F_{1x}\sin\phi, \\
\< P_{y}^{\mathrm{final}} \mathrm{sign}(\eta) \> & =-F_{1y}\cos\phi.
\end{align}
The feed-downs suppress the transverse local $\La$ polarization by factors of $F_{1x}/\overline{f}_{1x}=F_{1y}/\overline{f}_{1y}=0.93$.

Figure~\ref{fig:decayFz} shows the effect of the feed-down on the longitudinal local $\La$ polarization. The blue curve is the fit to the final $\La$ polarization by the following equation:
\begin{equation}
P_{z}^{\mathrm{final}}(\phi) = F_{z}\sin(2\phi).
\end{equation}
The effect of the feed-downs is suppression to the primordial polarization, and the corresponding factor is $F_{z}/f_{z}=0.87$.

From the above calculation, we can see that, in all cases, the feed-downs can reduce the $\La$ polarization by a factor of $\sim0.9$ but cannot flip the sign of the primordial polarization.

\subsection{Discussions} \label{sec3C}

\begin{figure}[t]
\centering \includegraphics[width=1\columnwidth]{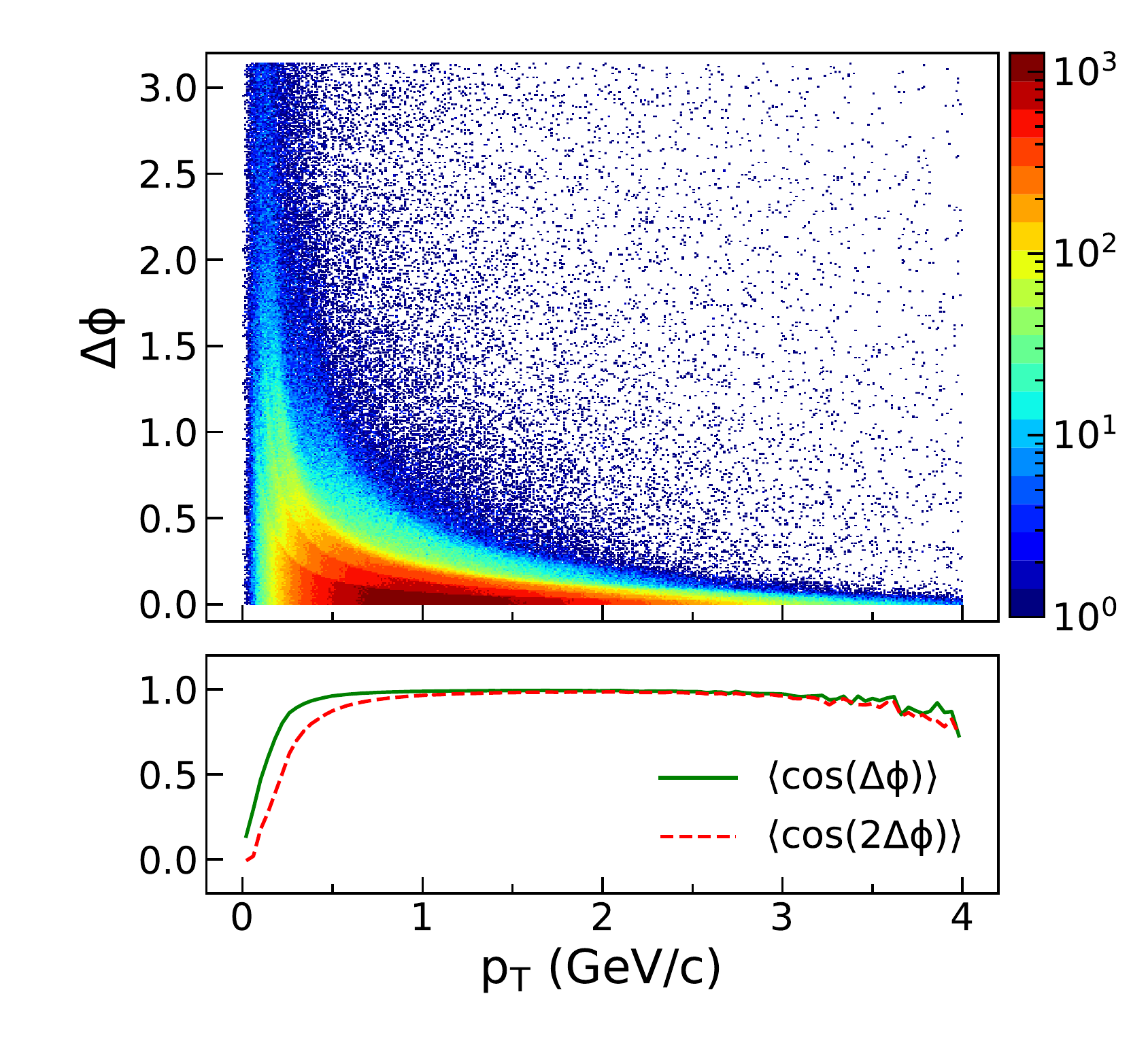}
\caption{The upper panel shows the distribution of final $\La$'s on the $\Delta\phi$-$\pT$ plane with $\Delta\phi$ as the azimuthal-angle difference between the parent and the daughter $\La$. The lower panel shows the averages of $\cos(\De\phi)$ (green line) and $\cos(2\De\phi)$ (red line) as functions of $\pT$. The simulation is for Au + Au collisions at $\sNN=200$ GeV.}
\label{fig:azimuthal_different}
\end{figure}

We present some discussions in order. (1) In the simulations, the input coefficients $f_0, \overline{f}_{1}, f_2$, and $f_z$ are chosen based on the available experimental data and the current model simulations. However, we have checked that, varying the values of the input coefficients has no significant change on our qualitative results. Furthermore, if the primordial $\La$ polarization is limited to be small $|\mathbf{P}^\mathrm{prim}_\La|<20\%$, which is the most likely case for realistic heavy ion collisions, our estimation for suppression factors of $F_0/f_0$, $F_{1}/\overline{f}_{1}$, $F_2/f_2$, and $F_z/f_z$ in the last subsection does not change significantly. (2) The main uncertainty for this paper comes from the choice of the primordial particle species. This is because the spin transfer law is different for different decay channels as shown in Table~\ref{tab:Polarization vector}. We note that our estimation for the suppression factor of the global polarization $F_0/f_0$ has a minor difference from the previous studies~\cite{Karpenko:2016jyx,Li:2017slc}. This is mainly because $\Xi^{-}$, $\Xi^{0}$, $\Xi^{*-}$, and $\Xi^{*0}$ were not included in those studies. (3) The $\phi$ angle appearing in Eqs.~(\ref{eq:P_x_primary})--(\ref{eq:P_z_primary}) is the primordial particle's azimuthal angle $\phi_{\mathrm{prim}}$, whereas the $\phi$ angle in other equations in the last subsection is the final $\La$'s azimuthal angle $\phi_{\mathrm{final}}$ which is different from $\phi_{\mathrm{prim}}$ in principle. However, our simulations indicate that these two should align well with each other. To understand this, we show the distribution of the final $\La$'s on the $\Delta\phi$-$\pT$ plane with $\Delta\phi$ as the difference between $\phi_{\mathrm{prim}}$ and $\phi_{\mathrm{final}}$ in Fig.~\ref{fig:azimuthal_different}. We can see that, except for the very low-$\pT$ ($\pT<0.5$ GeV/$c$) region, $\phi_{\mathrm{final}}$ aligns with $\phi_{\mathrm{prim}}$ well. Such a strong alignment can be characterized by the averages of $\cos(\De\phi)$ and $\cos(2\De\phi)$ shown in the lower panel of Fig.~\ref{fig:azimuthal_different}. Therefore, in the laboratory frame, the daughter $\La$ with $\pT>0.5$ GeV/$c$ flies almost along the same direction as its parent particle. As a consequence, the final $\La$ polarization inherits very similar azimuthal-angle dependence with that of the primordial particles.

\section{Summary and outlook} \label{sec4}

To summarize, we have studied systematically the effect of feed-downs from high-lying strange baryons on the $\La$ polarization. We have derived the angular distribution and polarization vector of the daughter particle for different decay channels including the strong decay $1/2^{\pm}\to 1/2^{+}0^{-}$ and $3/2^{\pm}\to 1/2^{+}0^{-}$, the weak decay $1/2\to 1/2\ 0$, and the EM decay $1/2^{+}\to 1/2^{+}1^{-}$, based on the framework of the helicity formalism.

We present a numerical computation of the feed-down effect on $\La$ polarization with appropriate input of the initial polarization and kinetic distribution of the primordial baryons. The high-lying baryons included in our simulation are listed in Table~\ref{tab:particle}. The numerical computation performed for Au + Au collisions at $\sNN=200$ GeV show that only about 21\% of the final $\La$s are primordial, and the others are produced by decays from other baryons. We find that the decays from $\Xi^0$ and $\Xi^-$ can lead to a radial $\La$ polarization opposite to the momentum direction of the produced $\La$. After a series of Monte Carlo simulations, we find that the feed-down contribution is not strong enough to flip the sign of the primordial polarization, although it suppresses the $\La$ polarization by a factor of $\sim0.9$ at $\sNN=200$ GeV. Therefore, we conclude that the feed-down effect does not solve the puzzle on the opposite azimuthal-angle dependence in the observed and predicted $\La$ polarization.

In the future, we will extend the current paper to other collision energies, especially to the energies covered by the Beam Energy Scan Program at the Brookhaven National Laboratory Relativistic Heavy Ion Collider. Besides, our paper can also provide guidance for the spin-polarization measurement of high-lying weak decay states, such as the $\Xi^-$ hyperon. The concrete simulation will be reported elsewhere.

{\it Note added.} During the preparation of this paper, we learned that Becattini {\it et al.}~have been working on the same subject~\cite{Becattini:2019ntv}. Their results bear some overlap with ours.

\begin{acknowledgments}

We thank F.~Becattini, G.~Cao, and S.~Choudhury for useful discussions. This work was supported by the NSFC through Grants No.~11535012, No.~11675041, and No.~11835002. X.-L.X. was also supported by the China Postdoctoral Science Foundation under Grant No.~2018M641909.

\end{acknowledgments}

\bibliographystyle{apsrev4-1}
\bibliography{ref}

\end{document}